\documentclass[sigconf]{acmart}

\def\BibTeX{{\rm B\kern-.05em{\sc i\kern-.025em b}\kern-.08emT\kern-.1667em\lower.7ex\hbox{E}\kern-.125emX}}

\usepackage[bottom]{footmisc}
\usepackage{booktabs} %
\usepackage{verbatim}
\usepackage[linesnumbered,algoruled,boxed,lined]{algorithm2e}
\usepackage{graphicx}
\usepackage{graphicx}
\usepackage{rotating}
\usepackage{graphicx}
\usepackage{caption}
\usepackage{cuted}
\usepackage{adjustbox}
\usepackage{enumitem,kantlipsum}
\usepackage{dsfont}
\usepackage{pgfplots}
\usepackage{amsmath}
\DeclareMathOperator*{\argmax}{\arg\!\min}

\usetikzlibrary{arrows}

\usepackage{amsthm}
\usepackage{arydshln}

\theoremstyle{definition}

\theoremstyle{remark}

\usepackage{subcaption}

\setcopyright{iw3c2w3}

\copyrightyear{2021}
\acmYear{2021}
\acmConference[WWW '21]{Proceedings of the Web Conference 2021}{April 19--23, 2021}{Ljubljana, Slovenia}
\acmBooktitle{Proceedings of the Web Conference 2021 (WWW '21), April 19--23, 2021, Ljubljana, Slovenia}
\acmPrice{}
\acmDOI{10.1145/3442381.3449970}
\acmISBN{978-1-4503-8312-7/21/04}

\begin{document}

\title[Random Walks with Erasure]{%
Random Walks with Erasure: Diversifying Personalized Recommendations on Social and Information Networks}

\author{Bibek Paudel}
\orcid{1234-5678-9012}
\affiliation{%
  \institution{Stanford University}
  \streetaddress{Stanford, USA}
  \city{Stanford, USA}
}
\email{bibekp@stanford.edu}

\author{Abraham Bernstein}
\affiliation{%
    \institution{University of Z\"urich}
  \streetaddress{University of Z\"urich}
  \city{Z\"urich, Switzerland}
}
\email{bernstein@ifi.uzh.ch}

\renewcommand{\shortauthors}{B. Paudel and A. Bernstein}

\begin{abstract}

Most existing personalization systems promote items that match a user's previous choices or those that are popular among similar users. This results in recommendations that are highly similar to the ones users are already exposed to, resulting in their isolation inside familiar but insulated information silos. 
In this context, we develop a novel recommendation framework with a goal of improving information diversity using a modified random walk exploration of the user-item graph.
We focus on the problem of political content recommendation, while addressing a general problem applicable to personalization tasks in other social and information networks.

For recommending political content on social networks, we first propose a new model to estimate the ideological positions for both users and the content they share, which is able to recover ideological positions with high accuracy. 
Based on these estimated positions, we generate diversified personalized recommendations using our new random-walk based recommendation algorithm. 
With experimental evaluations on large datasets of Twitter discussions, we show that our method based on \emph{random walks with erasure} is able to generate more ideologically diverse recommendations. 
Our approach does not depend on the availability of labels regarding the bias of users or content producers. 
With experiments on open benchmark datasets from other social and information networks, we also demonstrate the effectiveness of our method in recommending diverse long-tail items.
\end{abstract}




\maketitle

%!TEX root = bpaudel_recsys_2019.tex
%\vspace{-3mm}
\section{Introduction}
\label{sec:introduction}

\begin{comment}
\begin{figure}[!tbp]
  \begin{subfigure}[a]{1\linewidth}
  \vspace{1cm}
    \includegraphics[width=1\textwidth]{20_figs/bbc_example.png}
    %\vspace{0.1cm}
    \caption{\footnotesize News stories from the BBC shared by pro-remain (left) and pro-leave (right) users on Twitter.}
   \label{fig:bbc_example}
  \end{subfigure}
      
      \vspace{0.25cm}

  \begin{subfigure}[b]{1\linewidth}
    %\includegraphics[width=1\textwidth]{webconf_2021_RWE/20_figs/telegraph_example_1_blurred.png}
    \includegraphics[width=1\textwidth]{20_figs/telegraph_example_textonly.png}
    %\vspace{0.1cm}
    \caption{\footnotesize 
    News stories from The Daily Telegraph shared by pro-remain (left) and pro-leave (right) users on Twitter.}
   \label{fig:telegraph_example}
  \end{subfigure}
\caption{%\footnotesize 
News stories from the same outlet (BBC in (\protect\subref{fig:bbc_example}) and The Daily Telegraph in (\protect\subref{fig:telegraph_example})) that were shared by people with opposing viewpoints about the 2016 EU referendum (Brexit) held in the UK.}
\label{fig:bbc_and_telegraph_example}

%\vspace{-5mm}
\end{figure}
\end{comment}

Users often rely on recommender systems and Online Social Networks (OSNs) to select news and other information they consume~\cite{mitchell2015millennials,aljukhadar2012using}.
%Motivated by the goal of increasing sales revenue and click rates, these systems typically filter information by optimizing only (or mostly) on prediction accuracy.
%They promote items that match a user's previous choices or are popular among other similar users.
%As a result, they have been observed to result in familiar and popular recommendations, which isolate users in their existing information silos. 
%This has far-reaching impacts on social and democratic processes, as mass media and social networks heavily influence opinion formation and the level of civic engagement of citizens as well as political actors.
However, these algorithmic systems have also been criticized for increasing polarization and influencing political processes and events around the world.
%Therefore it, there is a need to investigate different strategies for recommending political content and other information to readers.
This makes it imperative to investigate better strategies for recommending personalized content to users.
%
\begin{comment}
The design of OSNs and Recommender Systems not only effect the attraction of their content-providers, but also the quality of information consumed by the wider public. 
%Automated filtering systems form an integral part of modern mass media, and their impact is presumed to be very widespread, including in the political and social processes that take place offline.
Automated filtering systems form an integral part of modern mass media, and their impact extends to political and social processes that take place offline.
Mass media influences the norms, opinions, and interactions that shape democratic engagement. 
%Such an impact begins early in a person's life~\cite{carpini2004mediating,boulianne2016online}.
Today, the Internet is the primary source of political news in many Western countries~\cite{mitchell2015millennials}, offering people an ever-growing pool of news sources.
This abundance of news sources has been found to lead to information overload and feelings of inability to cope with the news flow in people of all ages, ultimately resulting in ``news fatigue''~\cite{nytfatigue}. 
Young people have been found to be affected even more due to their information literacy levels~\cite{benselin2016information,wineburg2016evaluating}, and especially in light of their online news consumption~\cite{schmitt2017too}.
\end{comment}
%
High-quality as well as balanced news consumption is vital for a functioning democracy~\cite{muller2014comparing,mansbridge2012systemic,coronel2003role}. 
To address these needs, \emph{this paper introduces recommender algorithms that are designed with the goal of increasing the reader's exposure to diverse information.} % {\color{blue} and help them in filtering their continuous flow effectively}}.  {\color{red} maybe cut the blue part? } 

To \emph{promote the diversity of views}, we propose a random-walk based algorithm that can generate diverse as well as accurate recommendations. We introduce a modified random-walk exploration of the user-item feedback graph in which random-walk traversals to certain nodes are systematically erased in order to lower their importance with respect to the starting node of the random walk. We call this new approach \emph{random walk with erasure}.  
Our approach based on this modified random walk exploration provides a general mathematical and algorithmic framework for diversifying recommendations that can be used in various domains. Our evaluation, however, mainly focuses on the problem of political content recommendations on social media platforms (e.g., Twitter).

Towards this goal, we also propose a method that can identify the political leanings of both users and the news items they share on OSNs. We collected datasets of tweets on political discussions from various countries, related to events such as general elections or referendums. 
%Based on the principle of homophily and existing literature from political-science~\cite{poole1985spatial,clinton2004statistical}, we develop a method to identify the political leanings of users and news items discussed in social media.
We exploit the sharing behaviour of users on social media related to particular political events in order to estimate their ideological positions on a one-dimensional scale.
Based on such information, our recommendation approach can suggest news items to users that \textit{purposefully exposes them to different viewpoints and increases the diversity of their information ``diet.''}\footnote{Note that users can still choose which stories they want to read. As such, we do not \emph{impose} a news ``diet'' on them but provide a balanced suggestions.}

A common way to diversify content is by including viewpoints from different outlets, assuming that ideological positions of political elites and news outlets are fixed over long durations.
In highly contested political events, however, this approach is likely to suffer from a major problem:
a set of viewpoints from politicians or news sources belonging to different %parties with 
%different 
ideologies can still be homogeneous. %if it includes viewpoints of policians supporting the same campaign during events like the Brexit referendum. 
%supporting the same campaign during events like the Brexit referendum
%could still be homogeneous if it includes viewpoints from the member of the two parties supporting the same campaign during events like the Brexit referendum. 
%In this way, i
%It is possible for a set of viewpoints diversified in terms of its coverage of different parties, to be very homogeneous for a specific political agenda. % of the topic of interest for the users. 
%To solve this problem, 
The left-right classification of users or politicians based on their long-term behavior like speeches, voting habits or social media follower patterns is also likely to fall short.

Table~\ref{tab:bbc_telegraph_examples} shows two examples of how content from the same outlet were shared by groups of people with opposing political viewpoints about the 2016 Brexit referendum in the UK. 
Although The Daily Telegraph is known as a conservative British newspaper\footnote{https://yougov.co.uk/news/2017/03/07/how-left-or-right-wing-are-uks-newspapers/} that supported the Leave campaign,\footnote{https://www.telegraph.co.uk/politics/0/heres-where-britains-newspapers-stand-on-the-eu-referendum/} its report about the backtracking of a campaign promise by Leave campaigners (first example in the bottom row, Table~\ref{tab:bbc_telegraph_examples}) was popular among the supporters of the Remain campaign, while other pieces were popular among the supporters of the Leave campaign (second example in the bottom row, Table~\ref{tab:bbc_telegraph_examples}). 
The two opposing groups also shared articles from the BBC (top row in Table~\ref{tab:bbc_telegraph_examples}) differently: the first article was shared more by the Remain supporters and the second article was shared more by Leave supporters. 

\begin{comment}
Figure~\ref{fig:bbc_and_telegraph_example} shows two examples of how content from the same outlet were shared by groups of people with opposing political viewpoints about the 2016 Brexit referendum in the UK. Although The Daily Telegraph is known as a conservative British newspaper\footnote{https://yougov.co.uk/news/2017/03/07/how-left-or-right-wing-are-uks-newspapers/} that supported the Leave campaign,\footnote{https://www.telegraph.co.uk/politics/0/heres-where-britains-newspapers-stand-on-the-eu-referendum/} its report about the backtracking of a campaign promise by Leave campaigners (left side of Figure~\ref{fig:telegraph_example}) was popular among the supporters of the Remain campaign, while other pieces were popular among the supporters of the Leave campaign (right side of Figure~\ref{fig:telegraph_example}). The two opposing groups also shared articles from the BBC (Figure~\ref{fig:bbc_example}) differently: the article shown on the left was shared more by the Remain supporters and the one shown on the right was shared more by Leave supporters. 
\end{comment}

As a first step towards tackling this problem, \emph{we propose a novel approach to recommendation diversification that incorporates ideological positions about particular political events learned from social media signals}. To the best of our knowledge this is the first work to deal with this problem. Our proposed solution has two components: (i) learning ideological positions of users, political elites, and web content as well as (ii) using the ideological positions to diversify recommendation based on random walks with erasure and a diversification strategy that exploits weak ties in social networks.

\begin{table}
    \centering
    \footnotesize
    \begin{tabular}{p{8mm} p{70mm}}
        \toprule
        BBC & Brexit: Gibraltar in talks with Scotland to stay in EU (\href{https://web.archive.org/web/20171211005648/https://www.bbc.com/news/uk-politics-eu-referendum-36639770}{link}) \\
        \cline{2-2}
        & Second EU referendum petition investigated for fraud (\href{https://web.archive.org/web/20171228052309/https://www.bbc.com/news/uk-politics-eu-referendum-36634407}{link}) \\
        \midrule
        The &  Nigel Farage: £350 million pledge to fund the NHS was 'a mistake' (\href{https://web.archive.org/web/20171227185218/http://www.telegraph.co.uk/news/2016/06/24/nigel-farage-350-million-pledge-to-fund-the-nhs-was-a-mistake/}{link}) \\
        \cline{2-2}
        Telegraph & Britain remains a great country with a great future (\href{https://web.archive.org/web/20201125023449/https://www.telegraph.co.uk/opinion/2016/06/29/britain-remains-a-great-country-with-a-great-future/}{link})\\
         \bottomrule
    \end{tabular}
    \caption{\footnotesize 
    News stories from the same outlet (BBC in the top row and The Daily Telegraph in the bottom row) that were shared by people with opposing viewpoints about the 2016 EU referendum (Brexit) held in the UK.}
    \label{tab:bbc_telegraph_examples}
\vspace{-2mm}
\end{table}

In this work, we use the one-dimensional ideological positions (left-right) for users and political content. The key difference from other approaches is that we identify such positions for political elites, users, and individual content (rather than content outlets) depending on the sharing patterns on social networks during specific political events. Additionally, we also propose a novel and effective recommendation strategy based on ideological positions.
%The reason for identifying the political leaning of stories rather than news outlets (or publishers) is the following: the delineation of content publishers or political actors into strict ideological positions can be useful in many scenarios, but it hides the complex, multidimensional, and often contradictory nature of political debates. Some articles published in a newspaper categorized as being left-leaning may appeal to right-leaning audiences. Similarly, politicians having conservative views on national or economic issues may take a centrist stance on provincial or cultural issues.

With experimental evaluations on social network datasets, we show that our method is able to generate more politically diversified recommendations to the users without overly sacrificing accuracy. To show that random walk with erasure (RWE) is a general method that is also effective in other social and information networks, we also evaluate it on benchmark datasets from other domains like movie and restaurant recommendations. We find that RWE can recommend both highly accurate and diverse items to the users.

\textbf{In summary, our contributions in this paper are the following:}
%\begin{enumerate}
  %\item 
  (i) we describe a new \textit{method to estimate ideological positions of not only users and elites, but also web content} shared on social networks such as Twitter,
  %\item 
  (ii) we introduce \emph{random walk with erasure} (RWE), a novel modified random walk based exploration of bi-partite feedback graphs that is useful for diversifying recommendations,
  %\item 
  (iii) on datasets of social media discussions we show experimentally that our recommendation method based on RWE is able to \emph{diversify political content recommendations}---in many cases without a loss of accuracy, while state-of-the-art recommender systems generate recommendations that are less ideologically diverse, and
  %\item 
  (iv) on open benchmark datasets from other domains, we show experimentally that our algorithm can provide a \emph{general framework for diversified recommendations}.
%\end{enumerate}

The rest of the paper is structured as follows. %We first describe the challenges of political content recommendation and summarize our approach in Section~\ref{sec:strategies}. 
We review related work and describe the challenges of political content recommendation in Section~\ref{sec:related_work}. 
We introduce notations and background concepts in Section~\ref{sec:prelim}. Then we introduce our new method \texttt{random walks with erasure} in Section~\ref{sec:rw_erasure}, and two diversification strategies in Section~\ref{sec:strategies}. In Section~\ref{sec:ideology}, we describe our methods for the identification of political leanings 
%using (a) elite-endorsement, and (b) 
jointly using elite- and content-sharing signals. Finally, we describe our experimental setup and discuss the results of our evaluation in Section~\ref{sec:experiment}, and conclude in Section~\ref{sec:conlcusion}.

\section{Related Work}
\label{sec:related_work}
In this section, we review relevant literature from various fields.

\textbf{Recommendation Diversity.} Several previous works~\cite{nguyen2014exploring,munson2009sidelines,antikacioglu2017post,paudel2017updatable,castells2015novelty,ge2011placing,lathia2010temporal,zhang2012auralist,adomavicius2012improving,paudel2018loss,paudel2017fewer}
have investigated the issue of recommendation diversity. %
They diversify recommendations %
by exploiting topics and tags, post-processing of recommendations, promoting long-tail items, and so on. 
As a result, each of the proposed approaches provides a different kind of diversity to the users. 
Since the question of what constitutes a diverse recommendation does not have a clear answer, we think it is desirable to have multiple approaches and strategies for recommendation diversification.

\textbf{Ideology Detection from Social Media.} Methods that use social media behavior to estimate political leanings of users~\cite{conover2011predicting,barbera2013,colleoni2014echo} can be compared to the multidimensional scaling method famously known as DW-NOMINATE~\cite{poole1985spatial}, that measures ideology of parliamentarians by analyzing legislative 
voting behavior.
Some recent works approach the problem of recommendation diversity using ideological positions~\cite{garimella2017reducing,matakos2017measuring,aslay2018maximizing,musco2018minimizing,lahoti2018joint}, but they either rely on outlet specific positions, or do not provide a complete recommendation framework. 
We not only address the problem with outlet-specific positions, but also provide an end-to-end recommendation framework, with extensive evaluations with state-of-the-art methods.

\textbf{Political Content Diversity.} In context of political content, there are additional challenges regarding the question of recommendation diversity. Exposure to diverse viewpoints, and cross-cutting discussions between users of different viewpoints may help widen their perspective and can be desirable for a healthy democracy.
However, it is not enough to just diversify information without regard for several factors that influence opinion formation. %
Research has shown that exposure to diverse political viewpoints can also lead to further polarization~\cite{bail2018exposure}, especially in case of individuals who hold a strong viewpoint on a particular side of the debate~\cite{wojcieszak2009deliberation}. These behaviors have been explained by selective exposure theory and confirmation bias in social sciences. Other works have found that reinforcement of strong beliefs is weaker in individuals with moderate viewpoints, or in case of two-sided neutral debates and cross-cutting exposure~\cite{peterson_kagalwala_2021,karlsen2017echo,heatherly2017filtering}, as explained by social science concepts like moderation theory and cross-cutting discussions.

\textbf{Weak Ties.} There is also evidence from social network theory that \emph{weak ties are important for exposure to diverse information}~\cite{granovetter1977strength}. A recent study also shows that weak-tie discussion frequency is positively correlated to online political participation~\cite{valenzuela2011social}.

\textbf{Political Polarization on Social Networks.} In the context of political news, the role of OSNs and recommender systems in political polarization has recently become subjects of public concern.
This has led to an increased attention on studying the systemic bias of filtering algorithms and developing ways to correct them.
A study by Facebook\cite{bakshy2015} demonstrates how algorithmic filtering affects users' exposure to news in OSNs.
To get around the negative effects of information filtering, some news organizations have started to offer curated lists 
containing diverse viewpoints\footnote{For example: \url{https://www.theguardian.com/us-news/series/burst-your-bubble} 
and \url{http://graphics.wsj.com/blue-feed-red-feed/}}
in order to balance their readers' exposure.
Browser extensions have also been 
developed\footnote{For example: \url{https://www.escapeyourbubble.com/} 
and \url{http://politecho.org/}} 
to inject news links to ``burst the bubble.''

The challenge of diversifying recommendations can be seen as part of the research on \textbf{AI and machine-learning biases}~\cite{patro2020fairrec, arduini2020adversarial, heidari2019moral, zou2018ai}.
Variations of random-walks on networks have been used before to diversify rankings or improving predictions~\cite{backstrom2011supervised,zhu2007improving,paudel2017updatable}, but a general recommendation framework has been lacking.

\textbf{Scholars have argued that exposure to diverse viewpoints} helps broaden their acceptance, if not agreement, with such viewpoints.
They are deemed essential for promoting political tolerance and deliberative democracy. %
Insulated discussions among like-minded participants, in contrast, can breed excessive confidence, extremism, contempt for others, and also violence~\cite{sunstein2001echo}.
The exchange of ideas among people with politically dissimilar groups is known as \emph{cross-cutting discussions}.
Research in this field has shown that awareness of rationales for oppositional viewpoints increases with the exposure to disagreement, and that affective polarization is negatively related to involvement in cross-cutting discussions~\cite{peterson_kagalwala_2021,mutz2002cross}.
This implies that greater network diversity reduces polarization by facilitating cross-cutting discussions.

\section{Preliminaries}
\label{sec:prelim}
We begin by introducing the concepts and notations relevant for later sections of the paper.

\subsection{Random Walks on the Feedback Graph}
\label{sec:prelim_random_walks}

The user-item feedback dataset is a $m \times n$ matrix $\bf{A}$. The entries of the feedback matrix %
contain the feedback from user $u_i$ $(i\in1\ldots m)$ on item $i_j$ $(j \in1 \ldots n)$. In our case, we use implicit feedback, meaning that these entries are either $1$ for present or $0$ for missing feedback.

In this work, we also use the bi-partite graph representation $\mathcal{G}$ of the user-item feedback dataset.
We model $\mathcal{G}$ as unweighted and undirected graph (all edges have the same weight), but we could also generalize the definitions to a weighted version.
The adjacency matrix of the bi-partite user-item graph has the dimension $(m+n) \times (m+n)$ and is constructed as shown in~\eqref{eq:adj_matrix}:

\begin{align}
\label{eq:adj_matrix}
\textbf{A}^{\mathcal{G}} = \bigl[ \begin{smallmatrix}0 & \textbf{A} \\ 
								\textbf{A}^T & 0 \end{smallmatrix}
							\bigr]
\end{align}

The transition-probability matrix $\textbf{P}$ for $\textbf{A}^{\mathcal{G}}$ is obtained by row-normalizing its entries: 

\begin{align}
\label{eq:transition_matrix}
\textbf{P} = \textbf{D}^{-1} \textbf{A}^{\mathcal{G}}, \text{ where } \textbf{D}_{ii} = \sum \textbf{A}^{\mathcal{G}}_i.
\end{align} 

$\textbf{D}$ is the degree matrix which has the degree of the nodes of the graph in its diagonal elements.
The transition-probability matrix $\bf{P}$ has some interesting properties. Its entries $P_{ij}$ encode the probability of a random-walk starting at node $i$ arriving at node $j$ in one step. 
Every odd-power of $\bf{P}$ (e.g., $\bf{P^3}$) represents the transition probabilities for random walks starting at one of the user vertices and arriving at one of the item vertices. 
Consider a row-vector $v_s \in \{0,1\}^{(m+n)}$ all of whose entries are zero, except at index $s$. Then, the vector-matrix multiplication $v_s \bf{P}^3$ gives the transition probabilities for three-step random walks starting at node $s$. 

Instead of starting with an initial state probability of $1$, if the random walkers start with a mass of $e_s$, the index $s$ of $v_s$ contains $e_s$ instead of $1$, and the transition probabilities are obtained by vector-matrix multiplication as before:
$v_s\bf{P}^k$ ,
where $k$ is the number of steps in the random walk.

\subsection{One-dimensional Ideological Positions}
A seminal work about the estimation of ideological positions is by~\citeauthor{poole1985spatial}~\cite{poole1985spatial}, who used roll-call data from the United States Congress to recover the political positions of its members, called their ideal points. These approaches place the politicians on a latent dimension, which is usually a point in the one-dimensional left-right scale. Using this ideological dimension, a politician can be said to be left- or right-wing depending on whether her estimated position is towards the left or right of the center.

In this work, we estimate ideological positions for not only political elites, but also common users and the political content (URLs) they share on social media. We denote the ideal points of user $u$, elite $e$, and content $i$ by $\theta_u$, $\phi_e$, and $\psi_i$. By learning these positions jointly, we place them on a common ideological scale where comparisons can be made about their relative positions. For example, user $u_p$ and URL $i_q$ can be said to share similar political stance if their ideological positions are nearby, i.e., $\mid \theta_p - \psi_q \mid$ is small.

\section{Random Walks with Erasure}
\label{sec:rw_erasure}
In this section we define our new random-walk method, called random walk with erasure (RWE).
In RWE, we introduce variation on the usual random walk in the forms of erasures. 

At certain steps in the random walk, erasures cause a fraction of the mass reaching the destination vertices to be erased and sent back to the origin vertex. In other words, for a vertex that receives a mass of $p$ from a random walk, a portion $p \times q$ is erased, where $0 \leq q < 1$ is the amount of erasure. The remaining mass stays at the vertex and the erased mass is sent back to the vertex from which the random walk started. 

We can express this in probabilistic terms: \textbf{erasure probability} $q$ defines the probability with which the walk reaching a destination vertex is erased and sent back to the origin vertex. At the next iteration, instead of the usual mass of $1$, the walker starting at the origin vertex starts her walk with the mass accumulated from the erasures in the previous iteration. It is important to note that at each iteration the initial mass in the starting vertex gets smaller and is always less than $1$. In this way, RWE induces different random walk transition probabilities than the usual random walk. 

The intuition behind RWE is to allow different probability distributions than those induced by the degree distribution of the graph. 
This provides the flexibility to favor certain nodes during the random-walk exploration, based on their attributes, or similarity with the origin vertex.
We exploit this property of RWE to diversify recommendations by proposing two different strategies described below. First we start with a formal definition of RWE. %

\vspace{-2mm}
\subsection{Formal Definition}

RWE proceeds like a regular random walk except for two important differences.
The first involves a erasure-matrix $\bf{Q}$ which encodes the node-specific erasure probabilities.
The second is the erasure process itself. We now describe them in detail.

The amount by which the walks arriving a vertex are erased is not the same for all vertices%
\textemdash they differ for each pair of random walk origin and current vertex.
These quantities are encoded in $\textbf{Q} \in [0,1)^{(m+n)\times(m+n)}$.
The entries $\textbf{Q}_{ij}$ indicate the erasure probabilities from destination vertex $j$ to the origin vertex $i$. 

Note the \textbf{distinction between PageRank and RWE}: while the restart probability and the erasure probability in the two method seem similar, the crucial difference is that the erasure probability is different for each pair of vertices and the number of walks that get erased varies in each iteration, until convergence.

At each iteration of the random-walk, RWE proceeds as follows: (a) start regular random walks of odd number of steps $k$ from origin vertex $s$, and as specified by the transition probabilities $\textbf{P}$ in~\eqref{eq:transition_matrix}, (b) at the destination vertex $j$, with probability $\textbf{Q}_{ij}$, erase the walk; with probability $1- \textbf{Q}_{ij}$, do not erase the walk, and (c) at the second iteration, start new random walks from the origin vertex with the following probability
\vspace{-2mm}
\begin{align}
(\textbf{P}^k\circ \textbf{Q}) \mathds{1} \circ \mathbf{I}_{.,s} 
\end{align}
\vspace{-2mm}

Here $\mathds{1}$ is a $m + n$ dimensional vector with all ones, $\circ$ is the Hadamard product, and $\textbf{P}^k\circ \textbf{Q}$ encodes erasures from the previous iteration. 
Multiplication with $\mathds{1}$ sums up all the erasures arriving at each origin-vertex. %
Considering $s$ as the origin user-vertex, $\mathbf{I}_{.,s}$ is the $s^{th}$ column of $(m+n) \times (m+n)$ dimensional identity matrix and the final Hadamard product gives the initial state probability modified due to erasures.

This process is continued for sufficient number of iterations and finally the number of walks at the destination vertices that were not erased is used to estimate the probability of a $k$-step RWE starting at $i$ and reaching $j$ without being erased. 
We use this probability to score item-nodes for recommendation tasks.

\section{Diversification Strategies}
\label{sec:strategies}
The Erasure matrix $\textbf{Q}$ can be defined by the service providers according to their strategy for diversifying recommendations. 
In other words, the strategy determined by $\textbf{Q}$ can be defined to favor diverse items that would be less traversed by regular $k$-hop random walks. 

Note that the items in the final recommendation list are those in the local neighborhood (when $k$ is not large) of the user vertices; only the probability of traversal to those vertices are changed due to $\textbf{Q}$. 
\textbf{RWE diversifies the recommendations by promoting diverse items connected by weak links}, and is less likely to recommend items that are too dissimilar or unfamiliar to the users.

In this section, we present two examples of diversification strategies: one for long-tail and one for political content diversity.

\subsection{Long-tail diversity (RWE-$D$)}
To use RWE for promoting long-tail diversity, as in $RP^3_{\beta}$~\cite{paudel2017updatable}, one can can define erasure matrix $\textbf{Q}^D$ as given in~\eqref{eq:rwe_d}, where $\textbf{D}$ defined in~\eqref{eq:transition_matrix} is the diagonal matrix containing vertex degrees, and $\beta$ is a parameter that can be used to tune the erasure probabilities. This strategy depends only on the degree of item vertices and has the effect of preferring low-degree (long-tail) items.

\begin{align}
\label{eq:rwe_d}
  \textbf{Q}^D_{i,j} = 1 - \frac{1}{\textbf{D}_{j,j}^\beta}
\end{align}

\subsection{Bridging political viewpoints (RWE-$B$)}

\begin{figure}
\centering
\begin{tikzpicture}[x=1.5cm] %

	\draw[latex-latex] (-2.5,0) -- (2.5,0) ;
	\foreach \x in  {-2,-1,...,2} %
		\draw[shift={(\x,0)},color=black] (0pt,3pt) -- (0pt,-3pt);
	\foreach \x in  {-1.5,-0.5,...,1.5} %
		\draw[shift={(\x,0)},color=black] (0pt,1.5pt) -- (0pt,-1.5pt);
    \foreach \i in {-2,-1,...,2} %
      \draw (\i,0.1) -- + (0,-0.2) node[below] {$\i$};
    
    \fill[blue](-0.9,0) circle (0.6 mm);
    \draw (-0.9,0.1) -- + (0,0) node[above] {$u_1$};
  	\fill[red](0.8,0) circle (0.6 mm);
  	\draw (0.8,0.1) -- + (0,0) node[above] {$u_2$};
  	\fill[red](1.1,0) circle (0.6 mm);
  	\draw (1.1,0.1) -- + (0,0) node[above] {$u_3$};
  	\fill[red](1.4,0) circle (0.6 mm);
  	\draw (1.4,0.1) -- + (0,0) node[above] {$u_4$};
\end{tikzpicture}
\caption{\footnotesize{
An example showing four users $u_1 \ldots u_4$ with their ideological positions.}
}
\label{fig:line_number}
\vspace{-1mm}
\end{figure}
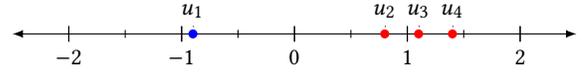

Before describing our strategies for diversifying political content recommendation, we illustrate example ideological positions of four users in Figure~\ref{fig:line_number}. Those towards the left %
in the ideological scale (e.g., $u_1$) are called left-leaning and those towards the right (e.g., $u_2, u_3, u_4$) are called right-leaning. These are relative comparisons with respect to other users, and we can also compare the distances and similarities between users based on their ideological positions: $u_3$ is more right leaning than $u_2$ but more left leaning than $u_4$, the distance between $u_1$ and $u_2$ is higher than between $u_2$ and $u_4$, and $u_2$ is more similar to both $u_3$ and $u_4$ than $u_1$ is.

In this work, we also identify ideological positions for political content (e.g., videos, news, social media posts, etc.) and elites. The ideological positions for user $u$, elite $e$, and content $i$ are $\theta_u, \phi_e, \ and \ \psi_i$, respectively.
Based on these positions, we define \textbf{similarity between a content-user or a elite-user pair} as: one minus the normalized absolute difference in their ideologiacal positions: $sim(i, u) = 1- (\mid \psi_i - \theta_u \mid)/(max_p-min_p)$ for content $i$ and user $u$, and $sim(e, u) = 1- (\mid \phi_e - \theta_u \mid)/(max_p-min_p)$ for elite $e$ and user $u$, where $[max_p, min_p]$ is the range of (all) ideological positions. It is symmetric and bounded between $[0,1]$

It is possible to transform the ideological positions (e.g., into $z$-scores), or to define similarity as a non-linear transformation of distance (e.g., using logarithmic transforms). We leave the study on the effect of such transformations for future work. %

As discussed in Section~\ref{sec:related_work}, %
the definition of diversity in the context of political content is far from clear. A diversification strategy that offers viewpoints radically different to a user's position is not likely to be appreciated by that user. Viewpoints that are different but not too far from the user's own ideological position---i.e., connected via weak links---can be expected to appeal more to the user than those that are at a greater distance. Such different viewpoints are likely to be reachable through others who are close to the user but in the opposite side of the political spectrum. We call them \emph{bridge users}, and in Figure~\ref{fig:line_number}, $u_2$ could be a bridge user between those on her right ($u_3 \ and \ u_4$) and those on her left ($u_1$). Similar notions apply in case of elites and content. \emph{Bridges are weak ties whose ideological positions are on the opposite side of the user's own position}.

Based on these motivations, we present the RWE strategy for bridging diverse political viewpoints and define the corresponding erasure matrix $\textbf{Q}^B$ for user-content pairs as:

\begin{align}
\textbf{Q}^B_{u,i} = 
\begin{cases}
sim(u,i), & \text{ if } $i$ \text{ is a bridge with respect to $u$}\\ 
\epsilon, & \text { otherwise} 
\end{cases},
\end{align}
where 
the values in $\textbf{Q}$ are less than $1$, and 
the parameter $\epsilon < 1$ is determined by the service provider. A high $\epsilon$ causes random walks reaching non-bridge elites or content to be erased at a higher rate. The erasure matrix $\textbf{Q}^B_{u,e}$ for user-elite pairs is defined similarly.
\section{Political Ideology Detection}
\label{sec:ideology}
In the previous sections, we discussed how we can use the ideological positions to find candidates for diversifying recommendations. Now we provide the details about our method to identify ideological positions for users, elites, and content. For this purpose, we consider two user-item feedback graphs: the elite-endorsement graph and the content-share graph. %
We construct these datasets from social media discussions around particular political events like elections, protests, or referendums. We use datasets of Twitter discussions, but the approach can be applied to similar datasets from any other social network. In the elite-endorsement graph, there are $m$ users $\mathcal{U}$ and $n_e$ elites $\mathcal{E}$. Elites are those individuals who are endorsed often, so it is possible for the same real-life person to be present both as a user and an elite in our dataset. The same users $\mathcal{U}$ and $n_i$ content-identifiers $\mathcal{I}$ constitute the content-share graph.

We treat retweets and content-sharing as acts of endorsements of elites and content by users with similar ideological positions. We consider any URL present in the tweets as a web-content and these URLs could refer to news, videos, pictures, or other social media posts. Similarly, we consider a retweet as an elite-endorsement. From these feedback graphs, we can construct two matrices similar to the feedback matrices defined in Section~\ref{sec:prelim_random_walks}: %
 $\textbf{R}$ of dimension $m\times n_e$ for the elite-endorsement graph and $\textbf{S}$ of dimension $m \times n_i$ for the content-share graph. The entries $\textbf{R}_{u,e}$ are $1$ if user $u$ has retweeted the elite $e$ and likewise entries in $\textbf{S}_{u,i}$ are $1$ if user $u$ has shared the content $i$. %
The remaining entries are zero.

\subsection{Using the elite-endorsement graph}
\label{sec:rt_ideology}
We assume a one-dimensional ideological space and want to recover the ideal points $\theta \in \mathbb{R}$ for users and $\phi \in \mathbb{R}$ for elites in this space. If a user and an elite share similar ideological positions, we assume the distance between them in this space to be low, and model these as quadratic utility functions similar to~\cite{clinton2004statistical,barbera2013}. With this assumption, $\textbf{R}_{u,e} = 1$ indicates that distance between $u$ and $e$ is small in this space. We model this in probabilistic terms, and state the probability of the user endorsing an elite using the logistic function:
\begin{align}
\label{eq:prob_retweet}
p(\textbf{R}_{u,e}=1 \mid \theta_u, \phi_e, \alpha_u, \beta_e) & = \frac{1}{exp(- \|\theta_u - \phi_e\|^2 + \alpha_u + \beta_e)}
\end{align}
where the terms $\alpha_{u}$ and $\beta_{e}$ are bias terms associated with $u$ and $e$, and account for the differences among users and elites respectively.

Using Bayesian inference, Bernoulli probability mass function, and under the assumption that all observed endorsements $\textbf{R}_{u,e}$ are independent, we get the following:

\begin{align}
\label{eq:lk_retweet}
p(\theta, \phi, \alpha, \beta \mid \textbf{R}) \propto \prod_{(u,e)\in \mathcal{U}\times\mathcal{E}} & p(\textbf{R}_{u,e}=1 \mid \theta_u, \phi_e, \alpha_u, \beta_e)^{a_{u,e}} \\
& (1-p(\textbf{R}_{u,e}=1 \mid \theta_u, \phi_e, \alpha_u, \beta_e))^{1-a_{u,e}} \nonumber
\end{align}

The parameter $a_{u,e}$ is used to assign confidence to the observed endorsement of $e$ by $u$, and it could be a function of the number of times $u$ has endorsed $e$. %

To simplify the notation, 
we write $\Pi_{u,e} = - \|\theta_u - \phi_e\|^2 + \alpha_u + \beta_e$.
After placing standard normal priors on $\theta$ and $\phi$, and taking the log of posterior, we get the following log-likelihood function with L-2 regularization terms: %

\begin{align}
\label{eq:lk_rt_final}
\argmax_{\theta, \phi, \alpha, \beta} \ p(\theta, \phi, \alpha, \beta \mid \textbf{R}) \propto & \sum_{(u,e)\in \mathcal{U}\times\mathcal{E}}
a_{u,e} (\Pi_{u,e}) - log (1 + exp(\Pi_{u,e})) \\
& - \frac{\lambda}{2}\|\theta\|_2 - \frac{\lambda}{2}\|\phi\|_2 \nonumber
\end{align}

\subsection{Using the elite-endorsement and content-share graph}
\label{sec:url_ideology}
To also identify the ideological positions of the URLs shared by users, we use the content-share graph and elite-endorsement graph together in a joint probabilistic framework. We assume that web-content shared by users have ideological positions $\psi\in\mathbb{R}$ in the same shared latent space described in Section~\ref{sec:rt_ideology}. In case of content, $\textbf{S}_{i,k}=1$ indicates that the distance between the ideological positions of $u_i$ and $i_k$ is small in this space. We also model the probability of a user sharing a web-content using logistic function:

\begin{align}
\label{eq:prob_url}
p(\textbf{S}_{u,i}=1 \mid \theta_u, \psi_i, \alpha_u, \gamma_i) & = \frac{1}{exp(- \|\theta_u - \psi_i\|^2 + \alpha_u + \gamma_i)}
\end{align}
where $\lambda_{i}$ is the bias term associated with $i$.

As before, using Bayesian inference and the assumption that all observed $\textbf{S}_{u,i}$'s are independent, we arrive the following expression:

\begin{align}
\label{eq:lk_url}
p(\theta, \psi, \alpha, \gamma \mid \textbf{S}) \propto \prod_{(u,i)\in \mathcal{U}\times\mathcal{I}} & p(\textbf{S}_{u,i}=1 \mid \theta_u, \psi_i, \alpha_u, \gamma_i)^{b_{u,i}} \\ 
& (1-p(\textbf{S}_{u,i}=1 \mid \theta_u, \psi_i, \alpha_u, \gamma_i))^{1-b_{u,i}} \nonumber
\end{align}

The parameter $b_{u,i}$ is used to assign confidence to the observed endorsement of $i$ by $u$, for example the number of endorsements. %

\begin{figure}[ht]
\begin{subfigure}{0.45\columnwidth}
    \centering
   \includegraphics[width=1.2in]
   {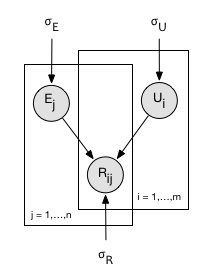}
   \label{fig:rt_graphical} 
\end{subfigure}
\begin{subfigure}{0.45\columnwidth}
    \centering
   \includegraphics[width=1.7in]
   {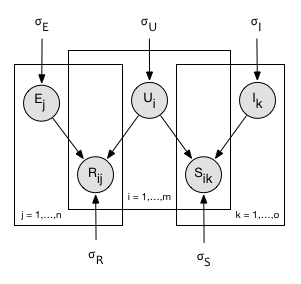}
   \label{fig:url_graphical}
\end{subfigure}
\vspace{-3mm}
\caption{\footnotesize{
Graphical models for the ideology detection in social networks using
endorsement of elites $E_j$ by users $U_i$ (left), and 
endorsements of elites and content $I_k$ by users (right), where the $\sigma_x$ are the priors for the elites, users, and items. 
}
}
\label{fig:graphical_models}
\vspace{-1mm}
\end{figure}

\paragraph{\textbf{Optimization.}} 
Instead of learning the ideologies in~\eqref{eq:prob_retweet} and~\eqref{eq:prob_url} separately, we formulate a \textbf{joint optimization} to learn all the ideological positions together.\footnote{More details, including source code are available online \url{https://github.com/bibekp/random_walks_erasure}} %
The reason for doing so is to align the positions learned by~\eqref{eq:prob_retweet} and~\eqref{eq:prob_url}. When there is not enough observed data in $\textbf{R}$ or $\textbf{S}$, one model is also expected to compensate for the lack of data in the other when learning $\theta$'s, $\phi$'s and $\psi$'s jointly. In other words, we use these two models to regularize each other such that they share the same latent dimension and learn relative distances between them in that shared space.
Graphical models for both approaches are shown in Figure~\ref{fig:graphical_models}.

After placing standard normal priors on $\psi$ as before, and adding contribution from~\eqref{eq:lk_url} as an additional regularizer on~\eqref{eq:lk_rt_final}, we get the following log-likelihood function for joint optimization, where $\mu$ trades-off the contribution from elite-endorsement graph:

\vspace{-2mm}

\begin{equation}
\begin{aligned}
\label{eq:lk_url_final}
\argmax_{\theta, \phi, \psi, \alpha, \beta, \gamma} \ & p(\theta, \phi, \alpha, \beta \mid \textbf{R}, \textbf{S}) \propto \\ 
& \mu \sum_{(u,e)\in \mathcal{U}\times\mathcal{E}}
a_{u,e} (\Pi_{u,e}) - log (1 + exp(\Pi_{u,e})) - \\ 
& \sum_{(u,i)\in \mathcal{U}\times\mathcal{I}} 
b_{u,e} (\Pi_{u,i}) - log (1 + exp(\Pi_{u,i})) \\ 
& - \frac{\lambda}{2}\|\theta\|_2 - \frac{\lambda}{2}\|\phi\|_2 - \frac{\lambda}{2}\|\psi\|_2
\end{aligned}
\end{equation}

The local maxima of~\eqref{eq:lk_rt_final} and~\eqref{eq:lk_url_final} can be found via a gradient-based optimization in which all but one parameter are fixed at each step and they are updated alternatively.

\section{Experiments}
\label{sec:experiment}
Now we describe the setup and the findings of our experiments.

\subsection{Dataset Collection and Properties}
\label{sec:datasets}

To evaluate the performance of our method and baselines on political content diversification, we crawled tweets during three political events and created these datasets: (a) \textbf{UK2016} from the 2016 EU referendum in the UK, (b) \textbf{US2016} from the 2016 US presidential elections, and (c) \textbf{DE2017} from the 2017 German federal elections. We used Twitter's search API to crawl the tweets containing any one of the common terms related to the online discussion about the events. 
We included the terms from~\cite{howard2016bots, howard2016bots_uk,kollanyi2017junk} and those appearing in Twitter's trending topic lists.
These search terms are shown in Tables~\ref{tab:brexit_tweet_search_terms},~\ref{tab:us_tweet_search_terms}, and ~\ref{tab:germany_tweet_search_terms}. 

For UK2016 and US2016 datasets, we gathered roughly equal number of tweets for each major campaign position (Remain and Leave) or presidential candidate (Donald Trump and Hillary Clinton). For DE2017 dataset, we included search terms representing each major political party and some general terms related to the election. 
For UK2016, we collected tweets from the day of the referendum to until about ten days later. For US2016, we collected tweets from a week before till the day of the election. For DE2017, we collected tweets from a day before the election to about ten days later.
To filter suspicious users and content, we removed tweets that were not retweeted more than 50 times, and also the tweets by users who had few followers or who did not tweet often. Note that this step may not filter out bots and automated accounts.

We created two user-item feedback graphs for each dataset: (a) elite-endorsement, and (b) content-endorsement. We treat each Twitter user who is retweeted more than five times as an elite and each URL that is included in more than five tweets as a web-content. For URLs, we apply some preprocessing steps like un-shortening and resolving re-directions.
The rows of both matrices denote users and columns denote elites (in $\textbf{R}_{u,e}$) and web-content (in $\textbf{S}_{u,i}$), respectively. 
Additionally, we evaluate RWE-$D$ on two benchmark datasets from recommender systems: Movielens-1M and Yelp-Restaurants.
Table~\ref{tab:tweet_dataset_statistics} shows the statistics of these %
datasets.

\begin{table}
    \centering
    \footnotesize
    \begin{tabular}{p{10mm} p{60mm}}
         & Search Terms  \\
        \toprule
        Pro-Remain & \#remain, \#voteremain, \#votein, \#strongerin, \#yestoeu, \#yes2eu, \#23leadnotleave, \#betteroffin, \#bremain \\
        \midrule
        Pro-Leave & \#leave, \#notoeu, \#voteleave, \#takecontrol, \#leaveeu, \#voteout, \#23no2eu, \#betteroffout, \#britainout \\
        \bottomrule
    \end{tabular}
    \caption{\footnotesize 
    Search terms used to create the UK2016 dataset.}
    \label{tab:brexit_tweet_search_terms}
\vspace{-2mm}
\end{table}

\begin{table}
    \centering
    \footnotesize
    \begin{tabular}{p{10mm} p{60mm}}
         & Search Terms  \\
        \toprule
        Pro-Trump & trump, donald trump, @realDonaldTrump \\%, trumppence, \#MAGA, americafirst, imwithyou, trumptrain, makeamericagreatagain, trumppence2016, neverhillary, deplorables, votetrump, lockherup, crookedhillary, latinosfortrump, realdonaldtrump, lawandorder, pepe, wakeupamerica, \#tcot, donaldtrump, \#AltRight\\
        \midrule
        Pro-Clinton & clinton, hillary clinton, @HillaryClinton \\%, imwithher, nevertrump, clintonkaine2016, clintonkaine, strongertogether, \#p2, \#ctl, lovetrumpshate, libcrib, voteblue2016, dumptrump, hillaryclinton, basketofdeplorables, factcheck, tntweeters, countrybeforeparty, uniteblue, blacklivesmatter, feelthebern, \#VoteDems\\
        \bottomrule
    \end{tabular}
    \caption{\footnotesize 
    Search terms used to create the US2016 dataset.}
    \label{tab:us_tweet_search_terms}
\vspace{-2mm}
\end{table}

\begin{table}
    \centering
    \footnotesize
    \begin{tabular}{p{70mm}}
      Search Terms  \\
        \toprule
        \#btw17, merkel, \#87Prozent, bundestagswahl17, \#linke, \#afd, \#cdu, \#spd, \#gruene, \#fdp, german election \\
        \bottomrule
    \end{tabular}
    \caption{\footnotesize 
    Search terms used to create the DE2017 dataset.}
    \label{tab:germany_tweet_search_terms}
\vspace{-2mm}
\end{table}

\begin{table}
    \centering
    \footnotesize
    \begin{tabular}{c c c c}
        \textbf{Name} & \textbf{\#Users} & \textbf{\#Items} & \textbf{\#Ratings} \\
        \toprule
        \multicolumn{4}{c} {\textbf{Elite-endorsement}} \\
        \toprule
        \textbf{UK-RT} & 10,547 & 2,439 & 71,310 \\
        \midrule
        \textbf{US-RT} & 7,913 & 968 & 71,310 \\
        \midrule
        \textbf{DE-RT} & 5,305 & 837 & 23,561 \\
        \bottomrule
        \multicolumn{4}{c} {\textbf{Content-endorsement}} \\
        \toprule
        \textbf{UK-URL} & 10,547 & 1,244 & 37,036 \\
        \midrule
        \textbf{US-URL} & 7,913 & 698 & 30,801 \\
        \midrule
        \textbf{DE-URL} & 5,305 & 424 & 8,212 \\
        \bottomrule
        \multicolumn{4}{c} {\textbf{Recommender System Benchmark}} \\
        \toprule
        \textbf{ML-1M} & 6,040 & 3,706 & 998,087 \\
        \midrule
        \textbf{Yelp} & 6,945 & 11,274 & 316,162 \\
        \bottomrule
    \end{tabular}
    \caption{\footnotesize 
    Statistics of elite-endorsement and content-endorsement datasets created from the three tweet-datasets, along with the benchmark recommender datasets.}
    \label{tab:tweet_dataset_statistics}
\vspace{-4mm}
\end{table}

\subsection{Political Ideology Detection}
\label{sec:experiments_ideology}
\begin{figure}[ht]
\centering
   \includegraphics[width=0.8\linewidth]{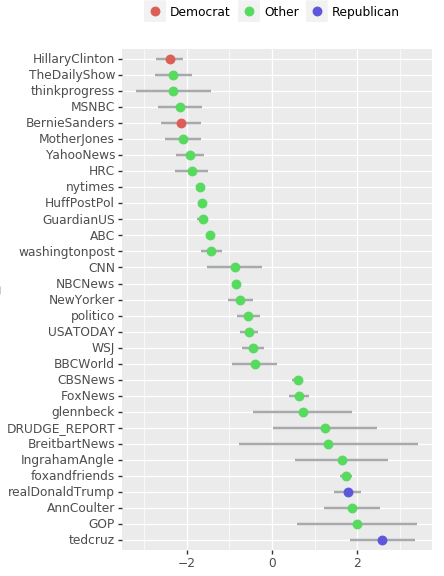}
      \caption{\footnotesize{
      Estimated ideological positions of political elites in 2016 US Presidential elections (US2016 dataset).}}
   \label{fig:us_positions} 
\end{figure}

\begin{figure}[ht]
    \centering
   \includegraphics[width=0.8\linewidth]{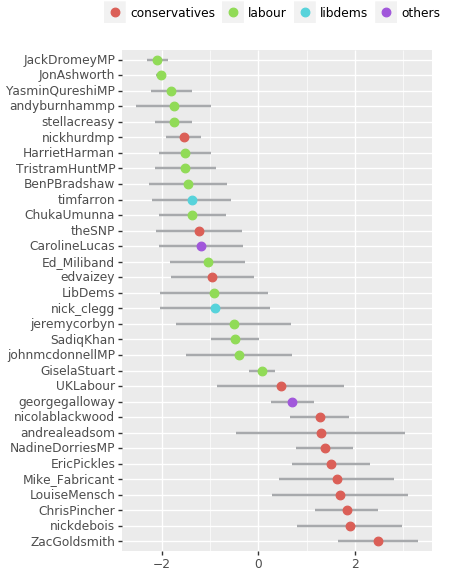}
   \caption{\footnotesize{
   Estimated ideological positions of political elites in 2016 EU Referendum in the UK (UK2016 dataset).}}
   \label{fig:uk_positions}
\end{figure}

\begin{figure}[ht]
\centering
   \includegraphics[width=0.8\linewidth]{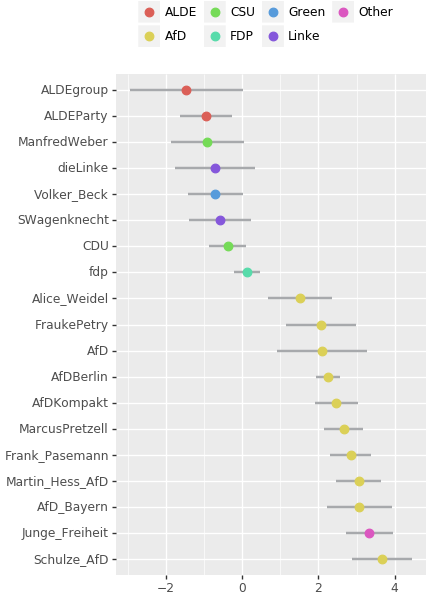}
      \caption{\footnotesize{
      Estimated ideological positions of political elites in 2017 German Federal elections (DE2017 dataset).}}
   \label{fig:germany_positions} 
\end{figure}

We evaluate our models introduced in Section~\ref{sec:ideology} in identifying political positions of political elites in the three datasets described in Section~\ref{sec:datasets}.
The ideological positions detected by the joint model in Section~\ref{sec:url_ideology} are given in Figure~\ref{fig:us_positions}, ~\ref{fig:uk_positions}, and ~\ref{fig:germany_positions} for US2016, UK2016, and DE2017 respectively. %

\begin{table}[ht]
    \centering
    \begin{tabular}{c c  c }
    \toprule
    Dataset & Elite endorsement & Joint learning from  
          elite-\\
          & only & and content-endorsement \\
        \toprule
        \bf{US2016} & 0.87 & \textbf{0.93} \\
        \bf{DE2017} & 0.56 & \textbf{0.64} \\
        \bf{UK2016} & 0.44 & \textbf{0.50} \\
        \bottomrule
    \end{tabular}
    \caption{\footnotesize
    {Pearson correlation coefficients for estimated ideological positions compared with baselines results~\cite{barbera2013}. Better result in each row is boldfaced. UK2016 shows a weak correlation; traditional left-right positions in the UK politics were not strongly reflected in the Brexit campaign. %
    }}
    \label{tab:ideology_compare_barbera}
    \vspace{-5mm}
\end{table}

The average and standard deviations from three separate runs are plotted as ideological positions on the x-axis. For Figure~\ref{fig:us_positions} and ~\ref{fig:uk_positions}, the elites on the y-axis are the Twitter users that are common to both our dataset and~\cite{barbera2013}. 
For Figure~\ref{fig:germany_positions}, the common Twitter users (with~\cite{barbera2013}) mostly included media personalities, so we manually chose the users close to each major party with the highest number of Twitter followers.

\paragraph{\textbf{Result I: Our method accurately identifies ideological positions from social network signals.}}
We compare the performance of two algorithms in Section~\ref{sec:ideology}: positions learned from elite endorsement only and positions learned jointly by using by elite-endorsement and content-endorsement. 
The comparisons with~\cite{barbera2013} are given in Table~\ref{tab:ideology_compare_barbera}. 
We see that for all datasets, positions learned jointly by using both social network signals perform better than those learned by using the endorsement signal only.
The correlation is especially strong for US2016 dataset but is weak for UK2016 dataset. 
The reason is that during the EU referendum of 2016, traditional left-right distances in the UK politics were not strongly reflected in the campaign endorsements by the political elites.

In all datasets we see a general separation of political elites into their left and right positions. For example, Democratic primary candidates Hillary Clinton and Bernie Sanders are located on the left, and Republican primary candidates Ted Cruz and Donald Trump are located on the right in Figure~\ref{fig:us_positions}. An interesting observation can be made in Figure~\ref{fig:uk_positions} where most Labour party leaders are located in the left and most Conservative party leaders are located in the right. However Conservative party leaders Nick Hurd and Ed Vaizy campaigned for the Remain campaign and our method has located them on the left. The Scottish National Party (SNP) also advocated for the Remain position. Labour MP Gisela Stuart campaigned for the Leave campaign, and has been located rightmost among all Labour politicians by out method. A list of MPs and their endorsements for the Leave and Remain campaigns can be found online.\footnote{~\url{https://www.bbc.com/news/uk-politics-eu-referendum-35616946}} 
In case of Figure~\ref{fig:germany_positions}, we can see that the traditional left-right alignment of German parties like Green, Linke, ALDE, CDU, FDP, and ALD is reflected on the detected ideological positions.

\subsection{Recommendation Baselines and Measures}

To choose baselines for comparing RWE, we referred to a recent work that analyzed several popular recommender algorithms~\cite{dacrema2019we,dacrema2021troubling}. 
Since RWE is based on graph exploration, we searched for baselines representing graph-based algorithms, as well as other methods that have shown to produce highly accurate and diverse recommendations. 
For political content diversification, we include the state-of-the art graph based method RP$^3_\beta$~\cite{paudel2017updatable}, which also deals with long-tail item diversity. 
Additionally, we include P$^3$~\cite{cooper2014random}, which achieves high accuracy but does not deal with diversity. 
Note that RP$^3_\beta$ outperforms the popular matrix factorization baseline BPRMF~\cite{rendle2012bpr}. 
For this reason, we include~\cite{johnson2014logistic}, which is a strong matrix-factorization baseline (MF). 
Finally, we also compare with item-based collaborative filtering (CF)~\cite{sarwar2001item}. 
For long-tail diversity, we compare RWE-$D$ against CF, P$^3$ and RP$^3_\beta$ since previous work has shown that they all outperform latent-factor models~\cite{paudel2017updatable}.

Among the several measures for evaluating the accuracy of a recommender system, we use the common ones: AUC, Mean Rank (MR), Hit-rate (HR), and Precision(P) at top-10. Higher values of these measures indicate better accuracy.

For measuring long-tail diversity, we borrow the measures Gini-Diversity (GiniD@20), Personalization (Pers@20), Surprisal (Surp@20), and Average Item Degree (AvgDeg@20) from the literature~\cite{adomavicius2012improving,paudel2017updatable}. Higher values of GiniD, Pers and Surp, as well as lower value of AvgDeg indicate better diversity.

For ideological diversity, we measure the average range of ideological positions in the top-k recommendations as $RecRange@k = \frac{1}{|U|} \sum_{u \in U} max(pos(R_u[1:k])) - min(pos(R_u[1:k]))$, where $R_u$ is the ranked list of recommendations for user $u$ and $pos(x)$ gives the ideological position of $x$. 

To evaluate the recommender algorithms, we divide the datasets into train-test splits in the following way: for each user with more than three interactions, we randomly select 30\% of the items 
into test-set. The remaining interactions are selected into train-set. We repeat this process three times, creating three independent test-train splits.
We run each algorithm three times (once on each test-train split) and report the average results. 

For RWE-based algorithms, we use a parameter $\nu$ to change the values of erasure matrix as $\textbf{Q}^{\circ\nu}$ (this is an element-wise operator and preserves the properties of the erasure matrix), and use $\epsilon = 0.9$. 
To search for the best parameters, we do a grid search on $\nu$, $\beta$, and number of nearest neighbors $k$ in case of RWE
, RP$^3_\beta$, and CF respectively. In case of MF, we do a grid search on the number of components $k$ and regularization constant. 
We choose the parameters that result in the best AUC and report the corresponding results for all eight datasets in Table~\ref{tab:summary_comparison}, %
where measures related to accuracy and long-tail diversity are in columns 2-5 and 6-9 respectively.

\subsection{Long-Tail Diversity}
\label{sec:experiments_longtail_recsys}

\paragraph{\textbf{Result II: RWE generates accurate and diverse long-tail recommendations.}}
We can see from Table~\ref{tab:summary_comparison} that on all datasets RWE is able to achieve best accuracy result (measured by AUC, Hit Rate, Precision, and Mean Rank). 
On ML-1M and Yelp datasets, where RWE-$D$'s erasure matrix is designed to promote long-tail diversity, it achieves similar or second-best results to RP$^3_\beta$ (measured by Gini, Average Degree, Personalization, and Surprisal). It shows that in addition to political content, RWE is also suitable for general recommendation tasks (e.g., movies, or restaurants).
In terms of long-tail diversity, on the Twitter-based datasets RWE does not perform so well because its erasure matrices are designed to promote diversity of political positions.
While CF generates better long-tail diversity, it does so at a high cost to accuracy.

\subsection{Ideological Diversity}
\label{sec:experiments_ideological_recsys}

\paragraph{\textbf{Result III: RWE generates ideologically diverse recommendations.}}
\label{par:exp_pol_div}
In case of Twitter-based datasets for political content, we lack measures that comprehensively capture the recommendation diversity. In this section, we use four methods: (i) average range of ideological positions in the top-10 recommendations $RecRange@10$ (ii) visual comparison of ideological distribution of items in top-k recommendations, (iii) Kolmogorov-Smirnoff statistic to quantify the difference in distributions of political ideology in top-k recommendations, and (iv) new measures to numerically and visually inspect the ideological diversity for users across the spectrum. The first three results are presented below and the fourth can be found in Appendix~\ref{sec:appendix_results}.
In all cases we compare RWE-$B$ against the algorithms which are most competitive in terms of accuracy (P$^3$, and RP$^3_\beta$). 
Unless otherwise specified, the parameters corresponding to the best accuracy measure in Table~\ref{tab:summary_comparison} are used.
 
In Table~\ref{tab:avgspread_diff_algorithms}, we list the $RecRange@10$ values for the algorithms which are most competitive in terms of accuracy. We can observe that RWE-$B$ outperforms other methods in all but one dataset.

\begin{table}[]
    \centering
    \begin{tabular}{ r  l  l  l }
    \textbf{Dataset} & \textbf{RWE-$B$} & \textbf{RP$^3_\beta$} & \textbf{P3} \\
    \toprule
    \bf{US-RT} & \textbf{1.71*} & 1.61 & 1.43 \\
    \bf{US-URL} & \textbf{2.15***} & 1.73 & 2.0 \\
    \bf{UK-RT} & \textbf{2.83***} & 2.45 & 2.58 \\
    \bf{UK-URL} & \textbf{2.02***} & 1.87 & 1.93 \\
    \bf{DE-RT} & \textbf{3.14***} & 2.61 & 2.98 \\
    \bf{DE-URL} & 1.53 & \textbf{1.69***} & 1.47 \\
    \bottomrule
    \end{tabular}
    \caption{\footnotesize 
Average spread of ideological positions in the top-10 recommendations ($RecRange@10$) by different algorithms in three independent runs on each dataset. Best results are boldfaced and statistically significant difference with the second best results are indicated with asterisks (one tailed Welch's t-test, * p-value < 0.05, *** p-value < 0.001). RWE produces more iedologically diverse recommendations on all but the smallest dataset.}
\label{tab:avgspread_diff_algorithms}
\vspace{-6mm}
\end{table}

\begin{figure}[!tbp]
  
  \begin{subfigure}[a]{1\linewidth}
  \centering
        \includegraphics[height=1.65in]{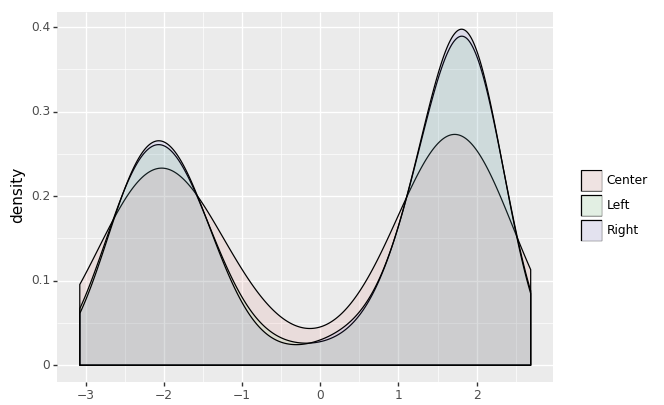}
        \caption{\footnotesize{Item-based Collaborative Filtering}}
        \label{fig:density_cf}
  \end{subfigure}
  \begin{subfigure}[c]{1\linewidth}
    \centering
        \includegraphics[height=1.65in]{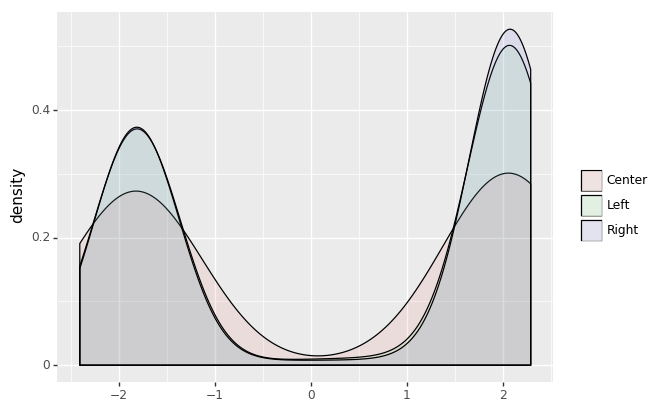}
        \caption{\footnotesize{Three-hop random walk ($P^3$)}}
        \label{fig:density_p3}
  \end{subfigure}
  \begin{subfigure}[d]{1\linewidth}
    \centering
        \includegraphics[height=1.65in]{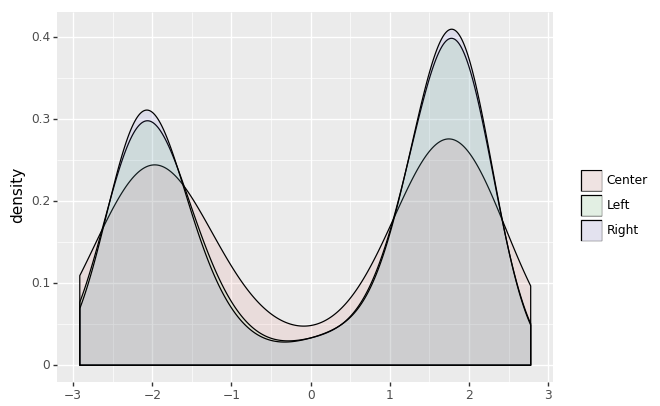}
        \caption{\footnotesize{Long-tail diversifying three-hop graph random walk ($RP^3_{\beta}$)}}
        \label{fig:density_rp3}
  \end{subfigure}
  
  \begin{subfigure}[f]{1\linewidth}
    \centering
        \includegraphics[height=1.65in]{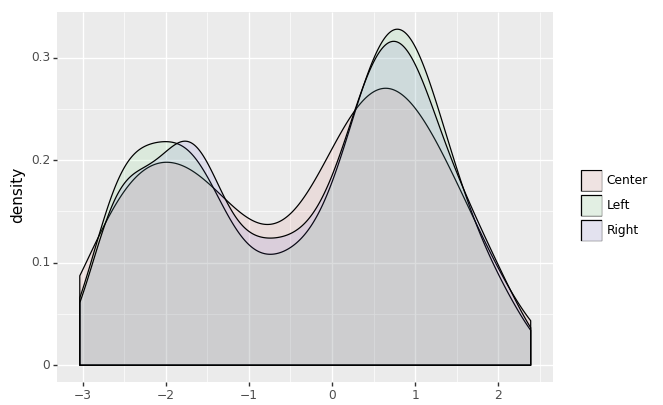}
        \caption{\footnotesize{Random walk with erasure (Bridging strategy) (RWE-$B$)}}
        \label{fig:density_rwemod}
  \end{subfigure}

\caption{\footnotesize 
Distribution of ideological positions in top-10 recommendations for left-, right-, and center-leaning users by the most competitive algorithms in the US-RT dataset. The colors indicate the ideological position of the users for whom the recommendations are generated and values in x-axis represent ideological positions of recommended items. RWE is able to generate recommendations that are more balanced and less polarized (more items from the middle), whereas other algorithms recommend more items from the extremes and less from the middle of the spectrum for all types of users. The Kolmogorov-Smirnov statistic was used to determine whether the recommendations from RWE were statistically different from those of other algorithms.}
\label{fig:density_plots_diff_algorithms}
\vspace{-5mm}
\end{figure}

To illustrate the diversification of recommendations by RWE-$B$, we compare the recommendations of the most accurate baselines with our approach on the US-RT dataset (results for other datasets are similar and omitted for space reasons). In this example, the task is to recommend elites to users (elites are \emph{items} in the traditional recommender system terminology). First, we classify users with ideological positions less than $-0.5$ as \emph{Left-leaning}, with positions greater than $0.5$ as \emph{Right-leaning}, and the rest as \emph{Center-leaning}. Then, we compare the distribution of ideological positions of items in the top-10 recommendations for all three types of users.
The results are shown in Figure~\ref{fig:density_plots_diff_algorithms}.

All the baselines shown in the figure achieve good accuracy, but they do so by recommending different kinds of political content. On the x-axis of these plots are the average ideological positions in the top-10 recommendations (elites), and colors of the density plots indicate the political ideology of the users for whom the recommendations are generated. The first three plots are for baseline algorithms: collaborative filtering (CF), three-hop random walk ($P^3$), and $RP^3_{\beta}$, which is a diverse recommender system based on $P^3$. The bottom plot is obtained from the recommendations of RWE-$B$, which uses bridging diversification strategy. We can see that the baselines' recommendations are more polarized, i.e., there are very few recommended items from the middle of the political spectrum, and most recommendations are either strongly left-leaning, or strongly right-leaning. 
\emph{Although biases in recommender systems have been studied before, they have not been examined before in the context of political content.}

In contrast, our algorithm is able to recommend more items from the middle of the spectrum. Similarly, the content is less polarized. 
As already discussed, even though it is not clear what constitutes a good diversification strategy,
RWE-$B$ is flexible enough to allow different diversification strategies to be plugged in to the system. Our strategy in Figure~\ref{fig:density_plots_diff_algorithms} shows that it is able to recommend more items that are both (i) dissimilar to the users' ideological positions and (ii) are not too far from the center of the spectrum.

We also used ~\textbf{Kolmogorov-Smirnov statistic} with the null hypothesis that the distribution of political ideology in the top-10 recommendations generated by RWE-$B$ is similar to those of baseline algorithms. With very high probability (p value << 0.0001), we are able to reject the null hypothesis (against all baselines).

Additional results demonstrating the higher ideological diversity of RWE-$B$'s recommendations are presented in Appendix~\ref{sec:appendix_results}.

\begin{table}[ht]
    \centering
    \tiny
    \begin{tabular}{ r || c | c | c | c || c | c | c | c }
    Model & AUC & HR@ & P@ & MR & Gini@ & AvgDeg@ & Pers@ & Surp@ \\
          &     & 10 & 10 &    & 20     & 20      & 20    & 20 \\ 
    \hline
    \multicolumn{9}{c} {\textbf{UK-RT}} \\
    \hline
    RWE-$B$ ($\nu=$0.1) & \bf{0.90} & \bf{0.22} & \bf{0.09} & 238.89 & 0.10 & 298.41 & 0.78 & 5.66 \\
    \hdashline
    RP$^3_\beta$ ($\beta=0.5$) & 0.89 & 0.13 & 0.06 & 257.49 & 0.49 & 119.85 & 0.96 & 8.24 \\
    CF ($k=50$) & 0.74 & 0.04 &  0.02  &  634.27 & \bf{0.52} & \bf{34.63} & \bf{0.98} & \bf{10.17} \\
    MF ($k=10$) & 0.85 & 0.10 & 0.05 & 377.0 & 0.16 & 392.36 & 0.53 & 4.92 \\
    P$^3$ & \bf{0.90} & \bf{0.22} & \bf{0.09} & \bf{234.69} & 0.10 & 299.53 & 0.78 & 5.67 \\
    \hline
    \multicolumn{9}{c} {\textbf{US-RT}} \\
    \hline
    RWE-$B$ ($\nu=0.1$) & \bf{0.92} & \bf{0.32} & \bf{0.14} & 76.92 & 0.09 & 437.33 & 0.62 & 4.34 \\
    \hdashline
    RP$^3_\beta$ ($\beta=0.5$) & 0.91 & 0.27 & 0.12 & 81.81 & 0.25 & 332.26 & 0.80 & 5.43 \\
    CF ($k=200$)        & 0.86 & 0.02 & 0.01 & 138.29 & \bf{0.54} & \bf{34.29} & \bf{0.96} & \bf{9.23} \\
    MF ($k=10$)        & 0.87 & 0.26 & 0.12 & 120.38 & 0.16 & 405.30 & 0.73 & 4.62 \\
    P$^3$                  & \bf{0.92} & \bf{0.32} & 0.13 & \bf{75.61} & 0.10 & 433.82 & 0.63 & 4.35 \\
    \hline
    \multicolumn{9}{c} {\textbf{DE-RT}} \\
    \hline
    RWE-$B$ ($\nu=0.1$)  & \bf{0.90} & \bf{0.37} & \bf{0.15} & \bf{85.68} & 0.20 & 142.33 & 0.79 & 4.81 \\
    \hdashline
    RP$^3_\beta$ ($\beta=0.5$) & 0.89 & 0.28 & 0.13 & 92.60 & 0.49 & 73.70 & 0.94 & 6.37 \\
    CF ($k=400$)        & 0.86 & 0.08 & 0.04 & 119.42 & \bf{0.61} & \bf{17.02} & \bf{0.96} & \bf{8.25} \\
    MF ($k=10$)        & 0.82 & 0.19 & 0.09 & 149.14 & 0.21 & 222.20 & 0.42 & 3.67 \\
    P$^3$                  & \bf{0.90} & \bf{0.37} & \bf{0.15} & 87.09 & 0.20 & 142.63 & 0.80 & 4.81 \\
    \hline
    \multicolumn{9}{c} {\textbf{UK-URL}} \\
    \hline
    RWE-$B$ ($\nu=0.1$)   & \bf{0.85} & \bf{0.32} & \bf{0.11} & 191.39 & 0.27 & 171.65 & 0.84 & 5.72  \\
    \hdashline
    RP$^3_\beta$ ($\beta=0.5$) & 0.84 & 0.23 & 0.09 & 197.00 & \bf{0.65} & 58.90 & \bf{0.97} & 7.46 \\
    CF ($k=400)$         & 0.82 & 0.13 & 0.05 & 227.70 & 0.58 & \bf{18.00} & \bf{0.97} & \bf{8.51} \\
    MF ($k=10$)          &  0.75 & 0.16 & 0.07 & 307.57 & 0.15 & 254.35 & 0.48 & 4.38 \\
    P$^3$                   & \bf{0.85} & \bf{0.32} & \bf{0.11} & \bf{187.69} & 0.27 & 170.53 & 0.84 & 5.73 \\
    \hline
    \multicolumn{9}{c} {\textbf{US-URL}} \\
    \hline
    RWE-$B$ ($\nu=5.0$)   & \bf{0.91} & \bf{0.37} & 0.14 & 63.79 & 0.15 & 260.89 & 0.68 & 4.25 \\
    \hdashline
    RP$^3_\beta$ ($\beta=0.5$)  & 0.90 & 0.29 & 0.12 & 72.16 & 0.39 & 167.02 & 0.87 & 5.66 \\
    CF ($k=400$)         & 0.86 & 0.08 & 0.03 & 98.88 & \bf{0.49} & \bf{35.04} & \bf{0.94} & \bf{7.94} \\
    MF ($k=10$)          & 0.84 & 0.25 & 0.12 & 113.62 & 0.19 & 336.31 & 0.46 & 3.64 \\
    P$^3$                   & \bf{0.91} & \bf{0.37} & \bf{0.15} & \bf{62.71} & 0.15 & 257.13 & 0.68 & 4.27 \\
    \hline
    \multicolumn{9}{c} {\textbf{DE-URL}} \\
    \hline
    RWE-$B$ ($\nu=5.0$)   & \bf{0.83} & 0.34 & 0.14 & 72.50 & 0.38 & 51.06 & 0.84 & 4.02 \\
    \hdashline
    RP$^3_\beta$ ($\beta=0.5$)  & 0.82 & 0.31 & 0.14 & 75.03 & 0.61 & 31.35 & 0.92 & 4.87 \\
    CF ($k=200$)         & 0.81 & 0.21 & 0.10 & 80.43 & \bf{0.69} & \bf{19.38} & \bf{0.94} & \bf{5.56} \\
    MF ($k=25$)          & 0.73 & 0.14 & 0.06 & 112.18 &  0.26 & 77.53 & 0.58 & 3.10 \\
    P$^3$                   & \bf{0.83} & \bf{0.36} & \bf{0.15} & \bf{70.97} & 0.39 & 51.12 & 0.84 & 4.01 \\
    \hline
    \multicolumn{9}{c} {\textbf{ML-1M}} \\
    \hline
    RWE-$D$ ($\nu=0.7$)         & \bf{0.92} & 0.12 & 0.06 & 291.82 & 0.08 & 1430.58 & 0.70 & 2.37 \\
    \hdashline
    RP$^3_\beta$ ($\beta=0.7$)  & \bf{0.92} & \bf{0.13} & \bf{0.07} & \bf{272.24} & 0.14 & 1140.37 & 0.86 & 2.95 \\
    CF ($k=400$)                & 0.91 & 0.06 & 0.03 & 331.90 & \bf{0.22} & \bf{523.34} & \bf{0.94} & \bf{3.90} \\
    P$^3$                          & 0.89 & 0.09 & 0.04 & 387.69 & 0.24 & 1656.34 & 0.49 & 1.89 \\
    \hline
    \multicolumn{9}{c} {\textbf{Yelp}} \\
    \hline
    RWE-$D$ ($\nu=0.5$)         & \bf{0.94} & \bf{0.07} & \bf{0.03} & \bf{624.30} & 0.24 & 103.92 & 0.96 & 6.96 \\
    \hdashline
    RP$^3_\beta$ ($\beta=0.5$)  & \bf{0.94} & 0.06 & \bf{0.03} & 633.39 & 0.34 & 76.77 & 0.98 & 7.56 \\
    CF ($k=400$)                & 0.92 & 0.02 & 0.01 & 892.16 & \bf{0.37} & \bf{18.35} & \bf{0.99} & \bf{9.94} \\
    P$^3$                          & \bf{0.94} & \bf{0.07} & \bf{0.03} & 641.76 & 0.12 & 161.14 & 0.87 & 5.75 \\
    \hline
    \end{tabular}
    \caption{\footnotesize{Comparison of RWE-based algorithms with other baselines on political content diversification (first six datasets) and long-tail diversity (last two datasets). The dashed lines separate our methods with the baselines and the chosen parameter for each method is given in parentheses. The first four columns represent measures related to accuracy, the next four columns represent measures related to long-tail diversity. Best results are boldfaced.}}
    \label{tab:summary_comparison}
\end{table}

\section{Conclusion, Limitations, And Future Work}
\label{sec:conlcusion}
In this paper, we described the problem of diversifying personalized recommendations and its several challenges, especially in the context of political content. 
We proposed a novel approach to diversify recommendations in social and information networks and showed that it is able to generate both long-tail and ideologically diverse recommendations. 
Our recommendation algorithm is based on a novel random-walk based algorithm, called Random Walk with Erasure (RWE).
RWE iteratively samples the nodes in a user-item graph by preferring certain nodes over others, as specified by an erasure matrix. 
For ideological diversity, our approach consists of two parts: (a) detection of ideological positions of not just users and elites but also web-content by exploiting social media signals about important political debates, 
(b) diversification of recommendations using the detected ideological positions. 
To the best of our knowledge, this is the first work to present a framework for political content diversification and a joint learning of ideologies positions. 
To evaluate the performance of our algorithms, we compared the ideological positions of political elites during recent important events in three Western countries: Brexit referendum, US presidential elections, and German federal elections and showed that our joint learning framework can accurately detect ideological positions from social network signals. 
We also compared the recommendation performance of RWE with several baselines and showed that it is able to generate accurate and diverse recommendations.
Our work has the following assumptions and limitations.
First, we assume one-dimensional ideological positions, which is a simplification of real-world political debates.
Second, a proper measure for assessing political content diversification is still lacking and may be the subject of debate~\cite{Helberger2019}. %
Third, we need to test RWE in an real-world, interactive scenario.
Last, in the absence of bridge users, finding content that are both diverse and agreeable could be challenging. 
In practice, this could be addressed by introducing a threshold in our method to specify the maximum extent of diversification, or the minimum number of bridge-users needed to promote items.
We plan to explore these limitations and ways to diversify recommendations based on additional dimensions in future work.

Recent events have shown that the power of OSNs, online news outlets, and automated recommendations should not be underestimated. With our work, we hope to contribute to the endeavour of making machine learning algorithms and recommender systems support healthy debates to enable a better society.

\begin{acks}
      We would like to thank the Hasler Foundation for their generous support of some of the work presented here.
\end{acks}

\bibliographystyle{ACM-Reference-Format}
\bibliography{10_webconf2021_bibliography}

\appendix

\section{Appendix}
\label{sec:appendix}

\subsection{Additional Results}
\label{sec:appendix_results}

In addition to the results presented in Section~\ref{sec:experiments_ideological_recsys} of the main paper, here we introduce some new measures in order to help understand the difference in the ideological diversity of recommended items for users across the spectrum.
The top half of Table~\ref{tab:ideological_measures} contains measures for average ideological position of recommendations (Rec-pos) and training items (Train-pos), difference between Rec-pos and user's ideological position (User-shift) and between Rec-pos and Train-pos (Train-shift), and the range of recommended items' positions (Rec-range).
In the remainder of this section, we use a cutoff of $topk = 10$ for evaluating recommendation diversity.

\begin{figure}[t]%
  
  \begin{subfigure}{0.32\columnwidth}
    \centering
        \includegraphics[width=1.18in]{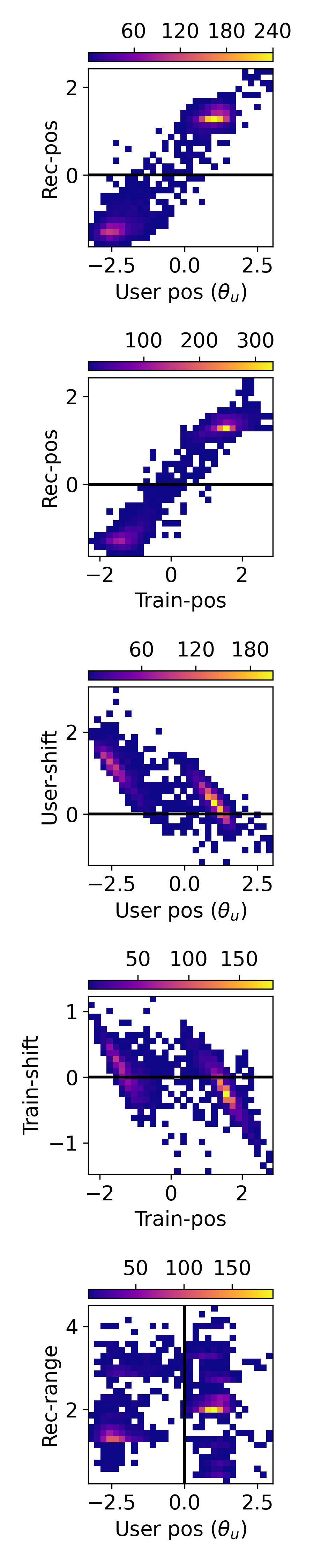}
        \caption{\footnotesize{$P^3$}}
        \label{fig:item_dist_p3}
  \end{subfigure}
  \begin{subfigure}{0.32\columnwidth}
    \centering
        \includegraphics[width=1.18in]{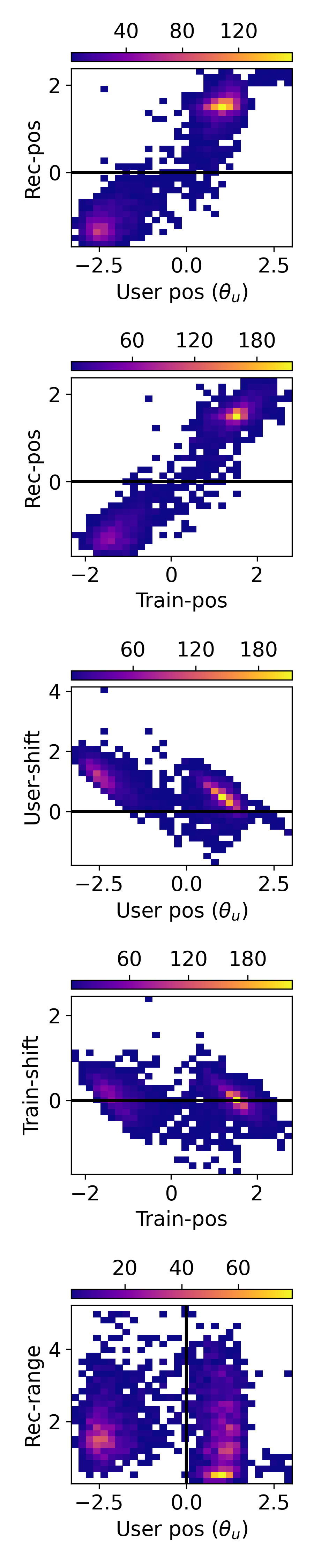}
        \caption{\footnotesize{$RP^3_{\beta}$}}
        \label{fig:item_dist_rp3}
  \end{subfigure}
  \begin{subfigure}{0.32\columnwidth}
    \centering
        \includegraphics[width=1.18in]{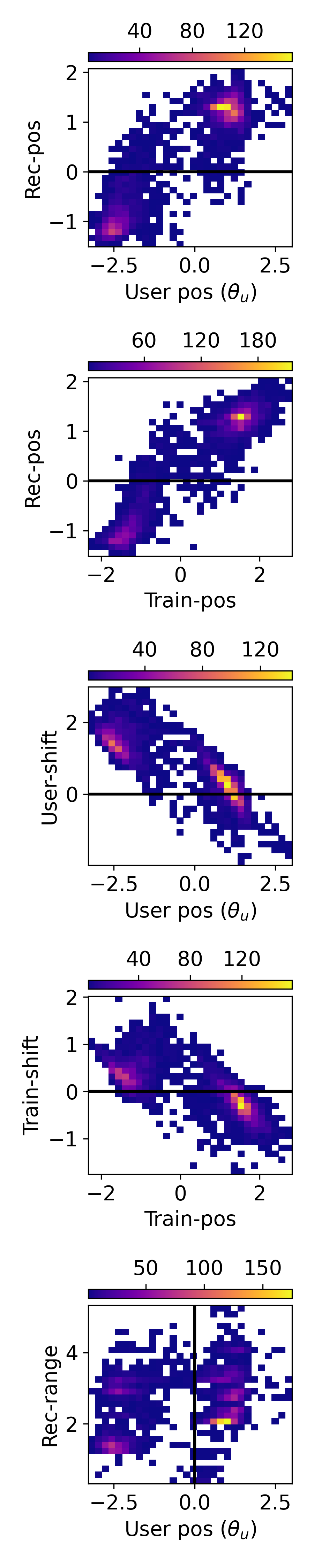}
        \caption{\footnotesize{RWE-$B$ ($\nu = 0.5$)}}
        \label{fig:item_dist_rwe}
  \end{subfigure}
\caption{\footnotesize{
Distribution of different ideological diversity measures for all users and top-10 recommendations in the US-URL dataset. Compared to the most competitive algorithms, RWE-$B$ is able to (i) recommend items whose ideological positions is different from the user and her observed preferences, (ii) shift the recommendations towards more moderate positions from user and her observed preferences, (iii) recommend items that span a wider range of ideological positions for all users.
} 
}
\label{fig:user_item_dist_diff_algorithms}
\vspace{-3mm}
\end{figure}

We show these measures for $P^3$, $RP^3_\beta$, and RWE-$B$ on the US-URL dataset in Figure~\ref{fig:user_item_dist_diff_algorithms}.
The relations of $\theta_u$ and Train-pos with Rec-pos can be seen in the first two rows, and their relations with User-shift and Train-shift can be seen in the third and fourth rows.
From the first two rows, we see that RWE-$B$ recommends more items from the upper and lower quadrants to users in the left and right respectively, and items from the center to users throughout the spectrum.
From the third and fourth rows, we see that RWE-$B$ has bigger positive shift for users in the left and bigger negative shift for users in the right.
In comparison, other algorithms have considerably less shift for most users and items are concentrated in the user's own quadrant, showing that their recommendations are not diverse.
The fifth row shows the relationship between $\theta_u$ and Rec-range, where RWE-$B$ has fewer recommendations with a narrow range (bottom of the plot) for users throughout the spectrum.

\begin{table}[h]
    \centering
    \footnotesize
    \begin{tabular}{ l l p{28mm} }
    \textbf{Name} & \textbf{Measure} & \textbf{Description}\\
    \toprule
      & $J_u = pos(R_u[1:k]$ & Positions of top-k recs\\
      & $L_u = pos(T_u[:])$ & Positions of trains\\
     Rec-pos & $pos(R_u) = \frac{1}{k} \sum J_u$ & Avg position of top-k recs\\
     Train-pos & $pos(T_u) = \frac{1}{|T_u|} L_u$ & Avg position of trainins\\
     User-shift & $shift(R_u,u) = pos(R_u) - pos(u)$ & Shift of recs from user\\
     Train-shift & $shift(R_u,T_u) = pos(R_u) - pos(T_u)$ & Shift of recs from trains\\
     Rec-range & $range(R_u) = max(J_u) - min(J_u)$ & Ideological range of recs\\
     \midrule
     UW-Recs & $\theta_u \times pos(R_u)$ & Rec pos weighted by $\theta_u$\\
     UW-Shift & $\theta_u \times shift(R_u,u)$ & User shift weighted by $\theta_u$\\
     TW-Recs & $pos(T_u) \times pos(R_u)$ & Rec pos weighted by Train pos\\
     TW-Shift & $pos(T_u) \times shift(R_u, T_u)$ & Train shift weighted by Train pos\\
     UW-Range & $ abs(\theta_u) \times range(R_u)$ & Rec range weighted by $\theta_u$\\
    \bottomrule
    \hline
    \end{tabular}
    \caption{\footnotesize 
Measures for ideological diversity in recommendations (recs = recommended items, trains = training items, pos = ideological position, $\theta_u$ = user $u$'s ideological position, $R_u$ are items recommended for user $u$, $T_u$ are training items for user $u$, $abs(\cdot)$ is absolute value of $x$, k is the threshold for top-k measures).}
\label{tab:ideological_measures}
\vspace{-4mm}
\end{table}

\begin{table}[h]
    \centering
    \begin{tabular}{rll | lll}
\hline
\bf{Measure} & \bf{RP$^3_\beta$} &  \bf{P$^3$} & \bf{RWE-$B$} & \bf{RWE-$B$} & \bf{RWE-$B$} \\
 & & & ($\nu=5.0$) & ($\nu=2.0$) & ($\nu=0.5$) \\

\hline
\bf{UW-Recs} &    1.99 &   1.90 &                1.79* &                \bf{1.29**} &                1.51** \\
\bf{TW-Recs} &    1.95 &   1.83 &                1.73 &                \bf{1.47**} &                1.56* \\
\bf{UW-Shift} &    -0.84 &  -0.92 &               -1.04* &               \bf{-1.53**} &               -1.31** \\
\bf{TW-Shift} &    -0.11 &  -0.24 &               -0.33 &               \bf{-0.59*} &               -0.49* \\
\bf{UW-Range} & 2.60 &   2.76 &                3.13* &                \bf{3.82**} &                3.60** \\
\bf{AUC} & 0.90 &  0.90 &                \bf{0.91} &                0.90 &                0.90 \\
\bf{HR@10} & 0.29 &   0.35 &                \bf{0.36} &                0.35 &                0.34 \\
\hline
\end{tabular}

\caption{\footnotesize{
Average measures of ideological diversity and recommendation accuracy in the US-URL dataset across three independent cross-validated runs. Best results are boldfaced and statistically significant difference with the best non-RWE result is indicated with asterisks (one tailed Welch's t-test, * p-value < 0.05, ** p-value < 0.01). RWE-B ($\nu=5.0$) has the highest accuracy and all parametarizations of RWE-$B$ produce more diverse result with accuracy similar to non-RWE baselines. Increasing the amount of erasure ($\nu = 2.0$ and $0.5$) shows the trade-off between accuracy and diversity. Lower values in the first four rows and higher values in the bottom three rows indicate better results.}}
\label{tab:summary_id_diverse}
\end{table}

To summarize the above measures for all users in a dataset, we introduce five measures based on the weightings of the measures described above by user's positions or training items' average position.
First, let us revisit the two (proposed) desirable properties for diverse recommendations: (i) recommendations should lean towards the center relative to a user's ideological position and (ii) recommendations should span a wider range of ideological positions for users in the extremes of the spectrum.
The first condition implies that for users on the left-side (or right-side) of the spectrum, a bigger positive difference (or negative difference) between the recommended and reference positions indicate higher diversity.
Similarly, for users with higher absolute positions, a wider range of recommendations are desired. %
These summary measures are presented in the bottom half of Table~\ref{tab:ideological_measures}, where
smaller values of UW-Recs, UW-Shift, TW-Recs, and TW-Shift and larger values of UW-Range indicate more diverse recommendations.

In Table~\ref{tab:summary_id_diverse}, we compare the performance of different parametarizations of RWE-$B$ with the most competitive baselines, where the measures are the averages across all users in the US-URL dataset for three indepent cross-validated runs.
We observe that $RWE$ can generate more ideologically diverse recommendations (low values of UW-Recs, TW-Recs, UW-Shift, TW-Shift, and high values of UW-Range).
We also observe a trade-off between accuracy and diversity as we change the amount of erasure.
The performance of RWE-$B$ with $\nu = 5.0$ is the most accurate (in terms of AUC and HR), but the performance of RWE-$B$ with higher amounts of erasure are also statistically indistinguishable from non-RWE baselines in terms of accuracy, while being more diverse.

\end{document}